\documentclass[12pt]{article}
\usepackage{epsfig}

\def\beq{\begin{equation}}
\def\eeq#1{\label{#1}\end{equation}}
\def\eeqn{\end{equation}}

\def\beqa{\begin{eqnarray}}
\def\eeqa#1{\label{#1}\end{eqnarray}}
\def\eeqan{\end{eqnarray}}

\let\bar=\overbar

\def\Dslash{\not{\hbox{\kern-4pt $D$}}}
\def\dslash{\not{\hbox{\kern-2pt $\del$}}}

\def\msb{{\bar{\ssstyle M \kern -1pt S}}}

\usepackage{fancyhdr,graphicx}
\def\Title#1{\begin{center} {\Large {\bf #1} } \end{center}}

\begin{document}

\Title{Implications of PSR J1614-2230 for NJL hybrid star}

\bigskip\bigskip


\begin{raggedright}
{\it C\'esar Henrique Lenzi \index{Lenzi, C. H.}\\
Universidade Tecnol\'ogica Federal do Paran\'a - Campus Medianeira \\
85884-000 - Vila S\~ao Cristov\~ao\\
Medianeira, PR\\
Brazil\\
{\tt Email: cesarlenzi@utfpr.edu.br}}
\bigskip\bigskip
\end{raggedright}

\begin{raggedright}
{\it Germ\'an Lugones \index{Lugones, G.}\\
Universidade Federal do ABC - Campus Santo Andr\'e\\
 09210-580- Bangu\\
Santo Andr\'e, SP\\
Brazil\\
{\tt Email: german.lugones@ufabc.edu.br}}
\bigskip\bigskip
\end{raggedright}

\section{Introduction}

The recent determination of the mass of the pulsar PSR J1614-2230 with $1.97 \pm 0.04 M_\odot$ by \cite{Demorest}, 
renewed the discussions about the possibility of exotic matter being present at  the core of neutron stars.
Since the description of matter at densities beyond nuclear saturation is model dependent, several works have explored different
aspects of the fact that the maximum neutron star mass implied by any equation of state (EoS) must exceed the mass of PSR J1614-2230
\cite{Lenzi2012,Bonanno2012}

In this article we present an extensive study of hybrid star masses using several parametrizations of a relativistic mean-field hadronic
EoS together with a typical three-flavor NJL model with scalar, vector and 't Hooft interactions as realized in \cite{Lenzi2012}.

\section{The model}

To describe the quark matter phase we use the SU(3) NJL model with scalar-pseudoscalar, isoscalar-vector and
't Hooft six fermion interaction. The Lagrangian density of the model is:
\begin{eqnarray}
{\cal L}_Q & = & \bar\psi(i \gamma_\mu \partial^\mu - \hat{m})\psi  \nonumber \\
           & + &  g_s \sum_{a=0}^{8} [( \bar\psi \lambda^a \psi )^2  +  (\bar \psi i \gamma_5 \lambda^a \psi)^2 ]  \nonumber \\
           & - &  g_v \sum_{a=0}^{8}   [(\bar\psi \gamma_\mu  \lambda^a \psi)^2 +  (\bar\psi \gamma_5 \gamma_\mu \lambda^a\ \psi)^2 ] \nonumber \\
           & + &  g_t\{\det[ \bar\psi(1+\gamma_5)\psi]   +  \det[ \bar\psi(1-\gamma_5)\psi]\}, 
\label{eq2}
\end{eqnarray}
where $\psi = (u,d,s)$ denotes the quark fields,  $\lambda^a ( 0 \leq a \leq 8 )$ are  the U(3) flavour matrices, 
 $\hat{m} = \mbox{diag}({m}_{u},{m}_{d},{m}_{s})$ is the quark current mass, and $g_s$, $g_v$ and $g_t$ are coupling constants. The
mean-field thermodynamic potential density $\Omega$ for a given baryon chemical potential $\mu$ at $T = 0$, is given by
\begin{eqnarray}
\Omega  = & - & \eta N_c\sum_i\int_{k_{Fi}}^{\Lambda}{\frac{p^2\, dp}{2\pi^2}}\sqrt{p^2+M_i^2} + 2g_s\sum_i \langle \bar \psi \psi \rangle_i^2  \nonumber \\
   & - & 2 g_v \sum_i \langle \psi^\dagger \psi \rangle_i^2 + 4g_t\langle\bar u u \rangle \langle\bar d d \rangle \langle\bar s s \rangle \nonumber  \\ 
  &  - & \eta N_c\sum_i \mu_{i}{\int_0^{k_{Fi}}  \frac{p^2\, dp}{2\pi^2}}-\Omega_0, 
\label{eq3}
\end{eqnarray}
where the sum is over the quark flavor $(i = u, d, s)$, the constants $\eta = 2$ and $N_c = 3$  are the spin and color 
degeneracies, and  $\Lambda$ is a regularization ultraviolet cutoff to avoid divergences in the medium integrals.
The Fermi moment of the particle $i$ is given by $k_{Fi} = \theta(\mu^{\ast}_{i} - M_i)\sqrt( \mu^{\ast 2}_{i} - M_i^2)$,
where $\mu^\ast_{i}$ is the quark chemical potential modified by the vectorial interaction, i.e.
$\mu^{\ast}_{u,d,s} = \mu_{u,d,s} - 4 g_v \langle \psi^\dagger \psi \rangle_{u,d,s}$. 

The conventional procedure for fixing the $\Omega_0$ term in Eq. (\ref{eq3}) is to assume that the grand thermodynamic potential
$\Omega$ must vanish at zero $\mu$ and $T$.
Nevertheless, this prescription is no more than an arbitrary way to uniquely
determine the EoS of the NJL model without any further assumptions \cite{Schertler1999}.
In view of this,  \cite{Pagliara2008}
adopt a different strategy.
They fix a bag constant for the hadron-quark deconfinement to occur at the same chemical potential as the chiral phase transition.
This method leads to a significant change in the EoS with respect to the conventional procedure.
Differently, in \cite{Lenzi2012} we explore the above possibility of having chiral restoration and deconfinement 
occurring at different densities. To this end, we shall substitute $\Omega_0$ in Eq. (\ref{eq3}) by the new value
$\Omega_0 + \delta \Omega_0$, where $\delta \Omega_0$ is a free parameter:
\begin{eqnarray}
\Omega_0  \longrightarrow  \Omega_0 + \delta \Omega_0  \  \ \ \ \   \mathrm{in} \ \mathrm{Eq.} \ (\ref{eq3}) .
\label{eq4}
\end{eqnarray}
With this change, the thermodynamic potential $\Omega$ can be non-vanishing at zero $\mu$ and $T$, and the $\mu$ of the
deconfinement transition can be tuned. For more details, see \cite{Lenzi2012}.

\section{Results}


We have solved the Tolman-Oppenheimer-Volkoff  equations for spherically symmetric and static stars in order to investigate the
influence of $g_v$ and $\delta \Omega_0$ on the maximum mass of hybrid stars.
In Figure 1 we have represented the maximum mass of hybrid stars for different parametrizations of the NJL model 
(more deals see the label of Figure).
An interesting feature of Figure 1, is that large masses are situated on the right-upper
corner but stable configurations are located on the left-lower corner of the figure (or left side of the figure in the case of NL3).
This clearly illustrates the difficulty of obtaining stable hybrid stars with arbitrarily large masses. 
Concerning the effect of the hadronic model we see that stable hybrid stars have higher values of the maximum mass for the stiffer
hadronic EoS.

\begin{center}
\includegraphics[width = 0.6 \textwidth]{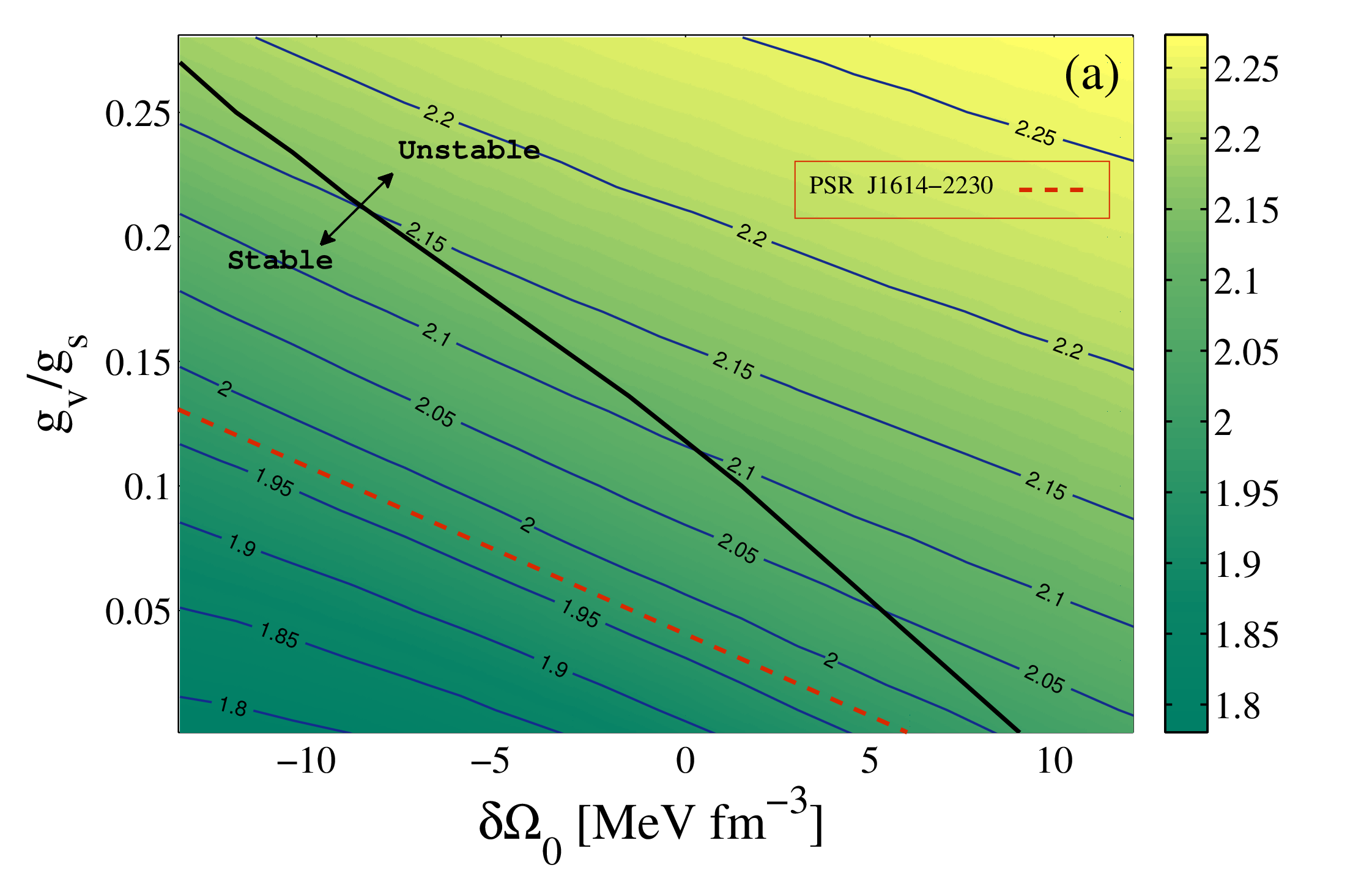}
\includegraphics[width = 0.6 \textwidth]{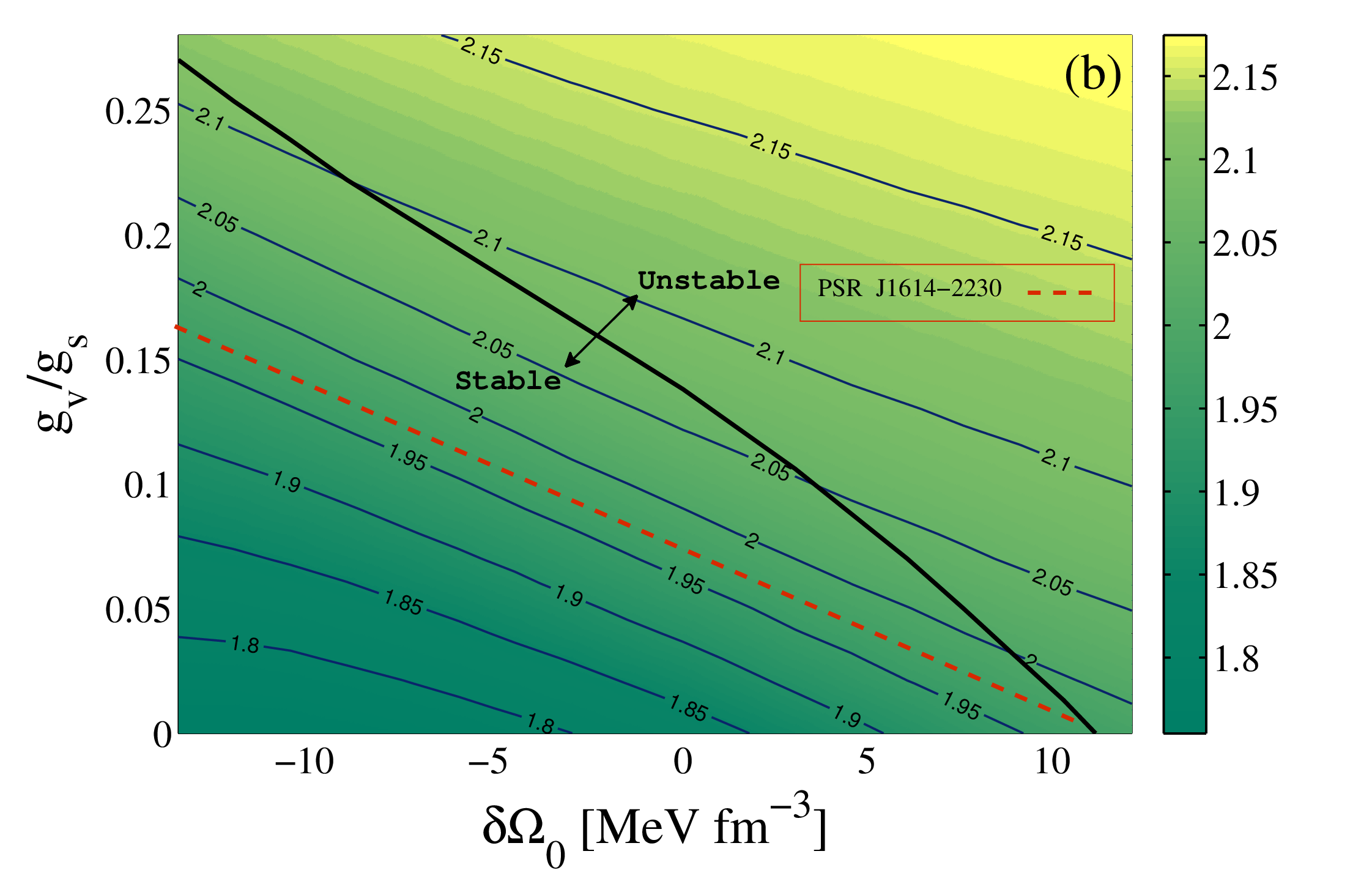}
\includegraphics[width = 0.6 \textwidth]{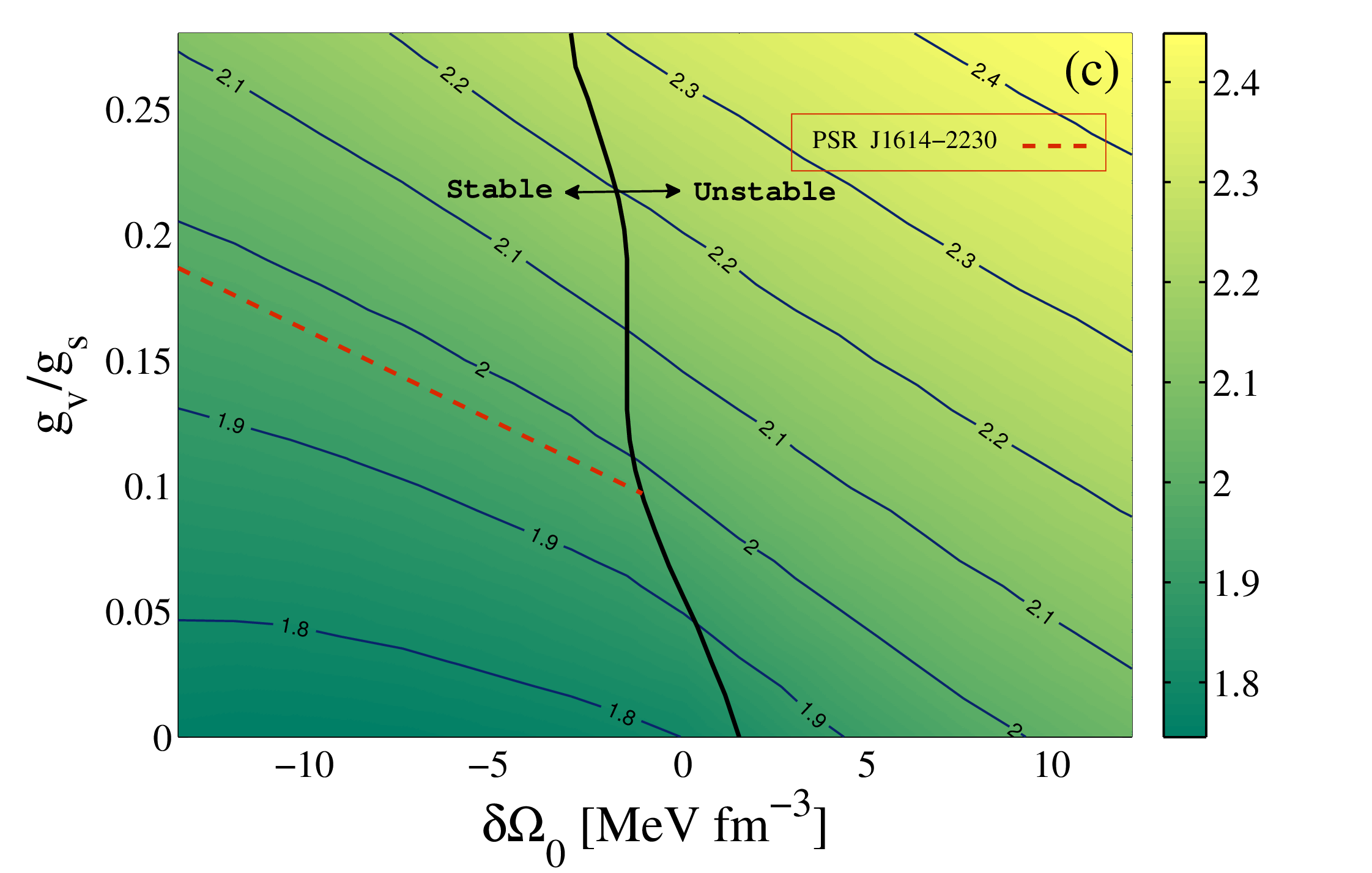}
\end{center}
{\footnotesize Figure 1: Background colors represent the maximum mass of hybrid stars for different parametrizations of the NJL model
(i.e. different values of $g_v$ and $\delta\Omega_0$).  In each panel we use a different hadronic EoS (without hyperons):
(a) GM1, (b) TM1 and (c) NL3. Notice that the color scale is different for each panel. The solid
contour lines indicate specific values of the maximum mass. The black solid line represents the boundary between parametrizations
that allow for stable hybrid stars and parametrizations that do not. The red dashed line indicates the value $1.97 {M}_{\odot}$ 
corresponding to the observed mass of PSR J1614-2230 \cite{Demorest}. The region between the red dashed line and the solid black
line allows to explain the mass of PSR J1614-2230. {\bf Font: Reference \cite{Lenzi2012}}.}

\ \ \\

The observed mass of PSR J1614-2230 can be explained by parameters within the large region located between the red dashed line and
the solid black line in each panel of Figure 1.  However, a hypothetical  future observation of a neutron star  with a
mass a  $\sim 10 \%$ larger than the mass of PSR J1614-2230 will be hard to explain within hybrid star models using the GM1 and
TM1 EoS and will require a very stiff hadronic model such as NL3.



\section{Conclusions}

In this work we performed a systematic study of hybrid star configurations using a relativistic mean-field hadronic EoS and
the NJL model for three-flavor quark matter. For the hadronic phase we used the stiff GM1 and TM1 parametrizations,
as well as the very stiff NL3 model. In the  NJL Lagrangian we included scalar, vector and 't Hooft interactions. 
The vector coupling constant $g_v$ was treated as a free parameter. We also considered that there is an arbitrary split between
the deconfinement and the chiral phase transitions. This split can be adjusted by a redefinition of the constant parameter 
$\Omega_0$  in the NJL thermodynamic potential.  
The effect of this redefinition is clear in Figure 1, where we show the maximum mass of static spherically
symmetric stars in the parameter space of $g_v$ and $\delta \Omega_0$. The larger masses are situated on the right-upper
corner of the diagram, where both $g_v$ and $\delta \Omega_0$ are larger. On the other hand, stable configurations are placed
on the opposite part of the diagram; i.e. on the left-lower corner for the GM1 and TM1 models and on the left side for the NL3 model.
As a consequence, stable configurations with a maximum mass compatible with PSR J1614-2230 are located halfway, specifically,
between the red dashed line and the solid black line of Figure 1. The effect of the hadronic model
(with nucleons only) is also clear from Figure 1, where we see that stable hybrid stars have higher values of the maximum mass for
the stiffer hadronic EoS. More details about this work can be seen in \cite{Lenzi2012}.

\end{document}